\documentclass[showpacs,preprintnumbers,amsmath,amssymb]{revtex4}
\usepackage{graphicx}
\usepackage{amssymb}

\begin{document}
\title{Secure quantum weak oblivious transfer against individual measurements}
\author{Guang Ping He}
\email{hegp@mail.sysu.edu.cn}
\affiliation{School of Physics and Engineering, Sun Yat-sen University, Guangzhou 510275,
China}

\begin{abstract}
In quantum weak oblivious transfer, Alice sends Bob two bits and Bob can
learn one of the bits at his choice. It was found that the security of such
a protocol is bounded by $2P_{Alice}^{\ast }+P_{Bob}^{\ast }\geq 2$, where $%
P_{Alice}^{\ast }$ is the probability with which Alice can guess Bob's
choice, and $P_{Bob}^{\ast }$\ is the probability with which Bob can guess
both of Alice's bits\ given that he learns one of the bits with certainty.
Here we propose a protocol and show that as long as Alice is restricted to
individual measurements, then both $P_{Alice}^{\ast }$\ and $P_{Bob}^{\ast }$%
\ can be made arbitrarily close to $1/2$, so that maximal violation of the
security bound can be reached. Even with some limited collective attacks,
the security bound can still be violated. Therefore, although our protocol
still cannot break the bound in principle when Alice has unlimited cheating
power, it is sufficient for achieving secure quantum weak oblivious transfer
in practice.
\end{abstract}



\maketitle



\section{Introduction}

Oblivious transfer (OT) \cite{qbc9,1-2OT} is known to be an essential
building block for two-party and multi-party protocols \cite{qi139}.
However, unconditionally secure OT was shown to be impossible even in
quantum cryptography, because the adversary can always cheat with the
so-called honest-but-curious attack \cite{qi499,qi677,qi725,qi797,qbc14}. To
evade the problem, the concept \textquotedblleft weak OT\textquotedblright\
was proposed recently \cite{qbc88}, in which the security goals of OT are
slightly modified, so that the honest-but-curious attack is no longer
considered a successful cheating. Even so, it was found that a security
bound exists for weak OT \cite{qbc88}, thus it cannot be unconditionally
secure either.

Nevertheless, we will point out below that the cheating strategy to weak OT
has its own limitation too. By making use of this limitation, we can build
quantum weak OT protocols which will violate the existing security bound
when the cheater is restricted to individual measurements as well as some
limited collective attacks. Therefore, while in principle the security bound
still applies to our protocols, in practice the attack will be very
difficult to be implemented.

Note that previously there was already a quantum OT protocol \cite{qi51}
which was considered secure against individual measurements \cite{qi138}.
However, the protocol calls for quantum bit commitment as a building block.
Thus its security is unreliable, as it is widely believed \cite{qi24,qi23}
that unconditionally secure quantum bit commitment does not exist.

In the next section, we will review the definitions of OT and weak OT, and
the existing security bound of the latter. Our protocol will be proposed in
Sect. III. Section IV is dedicated to the security analysis. We will show
why the security bound still applies in principle. On the other hand, we
will elaborate how to reach the maximal violation of the bound when only
individual measurements are allowed. It will also be shown that the bound
can still be violated for limited collective attacks. Some ideas on further
improvement of our protocol will be discussed in Sect. V. Finally, in Sect.
VI we summarize the result, and explain why it is important in practice to
find a secure protocol against individual measurements.

\section{Definitions and the security bound}

There are many variations of OT. The most well-known ones are all-or-nothing
OT \cite{qbc9} and one-out-of-two OT \cite{1-2OT}. Here we are interested in
the latter, which is defined as follows \cite{qi140}.

\bigskip

\textit{One-out-of-two Oblivious Transfer}

(i) Alice knows two bits $x_{0}$ and $x_{1}$.

(ii) Bob gets bit $x_{b}$ and not $x_{\bar{b}}$ with $Pr(b=0)=Pr(b=1)=1/2$.
(Here $\bar{b}$\ denotes the bit-compliment of $b$.)

(iii) Bob knows which of $x_{0}$ or $x_{1}$ he got.

(iv) Alice does not know which $x_{b}$ Bob got.

\bigskip

More rigorously, this definition indicates that a secure protocol should
guarantee that at the end of the process, Bob should get $x_{b}$ with
reliability $100\%$,\ i.e., the value he decoded matches Alice's actual
input with certainty. Meanwhile, the amount of information he gains on $x_{%
\bar{b}}$ should be arbitrarily close to zero, so that he has to guess $x_{%
\bar{b}}$ by himself, which results in a reliability $50\%$ for $x_{\bar{b}}$
since his guess stands a probability $1/2$ to be correct. However, as
pointed out in \cite{HeJPA}, in the literature there is the lack of a
self-consistent definition of OT specifically made for the quantum case.
This is because with quantum methods, it is possible that Bob may accept a
lower reliability of learning $x_{b}$, so that the reliability for $x_{\bar{b%
}}$ can be significantly raised. This is exactly what the honest-but-curious
attack \cite{qi499,qi677,qi725,qi797} achieves. In the above definition it
is vague whether such a result is considered as successful cheating, making
it hard to discuss the security of OT protocols in a precise way.

To mend the problem, weak OT is proposed \cite{qbc88}, with an improved
definition on Bob's cheating. Define the symbols

$P_{Alice}^{\ast }$: The maximum probability with which cheating-Alice can
get honest-Bob's choice bit $b$ and honest-Bob does not abort.

$P_{Bob}^{\ast }$: The maximum over $b\in \{0,1\}$ of the probability with
which cheating-Bob can get $x_{\bar{b}}$\ given that he gets $x_{b}$\ with
certainty and honest-Alice does not abort.

For every protocol there will always be $P_{Alice}^{\ast },P_{Bob}^{\ast
}\geq 1/2$, as a cheating party can do no worse than a random guess. Then
weak OT is defined as a kind of one-out-of-two OT which requires security
only against cheating-Bob who gets one of honest-Alice's bits with certainty.

Note that the names \textquotedblleft Alice\textquotedblright\ and
\textquotedblleft Bob\textquotedblright\ were used reversely in \cite{qbc88}%
, comparing with the literature on QOT \cite%
{qbc9,qi139,qi499,qi677,qi725,qi797,qbc14,qi51,qi138,qi140,HeJPA,HeQOT,qi135}%
. Here we follow the literature and use the names in the above way.

It was proven \cite{qbc88} that the optimal security bound for any quantum
weak OT protocol is%
\begin{equation}
2P_{Alice}^{\ast }+P_{Bob}^{\ast }\geq 2,
\end{equation}%
from which it follows that one of the two parties must be able to cheat with
probability at least $2/3$.

In brief, Ref. \cite{qbc88} obtained this bound with the following method.
First, consider Bob's cheating. Let $\rho _{b,x_{0},x_{1}}$ denote the
reduced state of his portion of the system. Since a weak OT protocol should
allow honest-Bob to learn $x_{b}$\ with certainty, there must be a
non-destructive measurement that enables him to do so without disturbing the
system. After Bob learned $x_{b}$\ with this measurement, his system will
still remain in the state $\rho _{b,x_{0},x_{1}}$. To gain some information
on the other bit $x_{\bar{b}}$, he performs the Helstrom measurement to
optimally distinguish the two states corresponding to $x_{\bar{b}}=0$\ and $%
x_{\bar{b}}=1$, respectively. Thus his cheating can be successful with
probability%
\begin{equation}
P_{Bob}^{\ast }\geq \frac{1}{2}+\frac{1}{8}\Delta ,  \label{qbc88-pb}
\end{equation}%
where%
\begin{equation}
\Delta \equiv \frac{1}{2}(\sum\limits_{x_{0}\in \{0,1\}}\left\Vert \rho
_{0,x_{0},0}-\rho _{0,x_{0},1}\right\Vert _{Tr}+\sum\limits_{x_{1}\in
\{0,1\}}\left\Vert \rho _{1,0,x_{1}}-\rho _{1,1,x_{1}}\right\Vert _{Tr}).
\end{equation}%
Secondly, consider Alice's cheating. She implements a uniform superposition
over $x_{0}$, $x_{1}$\ of honest strategies by introducing two additional
private qubits for storing the values of $x_{0}$ and $x_{1}$. Then she
applies the controlled-unitary operations%
\begin{equation}
controlled-U_{0}:\left\vert \psi _{1,1,x_{1}}\right\rangle \left\vert
1\right\rangle \left\vert x_{1}\right\rangle \rightarrow (I_{A}\otimes
U_{0,x_{1}})\left\vert \psi _{1,1,x_{1}}\right\rangle \left\vert
1\right\rangle \left\vert x_{1}\right\rangle
\end{equation}%
and%
\begin{equation}
controlled-U_{1}:\left\vert \psi _{0,x_{0},1}\right\rangle \left\vert
x_{0}\right\rangle \left\vert 1\right\rangle \rightarrow (I_{A}\otimes
U_{1,x_{0}})\left\vert \psi _{0,x_{0},1}\right\rangle \left\vert
x_{0}\right\rangle \left\vert 1\right\rangle
\end{equation}%
for $x_{1}\in \{0,1\}$\ and $x_{0}\in \{0,1\}$, respectively, where $%
U_{0,x_{1}}$ and $U_{1,x_{0}}$\ satisfy%
\begin{eqnarray}
F(\rho _{1,0,x_{1}},\rho _{1,1,x_{1}}) &=&\left\langle \psi
_{1,0,x_{1}}\right\vert (I_{A}\otimes U_{0,x_{1}})\left\vert \psi
_{1,1,x_{1}}\right\rangle ,  \nonumber \\
F(\rho _{0,x_{0},0},\rho _{0,x_{0},1}) &=&\left\langle \psi
_{0,x_{0},0}\right\vert (I_{A}\otimes U_{1,x_{0}})\left\vert \psi
_{0,x_{0},1}\right\rangle .
\end{eqnarray}%
Here $F(\rho ,\xi )\equiv \left\Vert \sqrt{\rho }\sqrt{\xi }\right\Vert
_{Tr} $\ is the fidelity between $\rho $\ and $\xi $. Then the successful
probability for her cheating was shown \cite{qbc88} to be%
\begin{equation}
P_{Alice}^{\ast }\geq \frac{1}{2}+\frac{1}{16}F,  \label{qbc88-pa}
\end{equation}%
where%
\begin{equation}
F\equiv \sum\limits_{x_{0}\in \{0,1\}}F(\rho _{0,x_{0},0},\rho
_{0,x_{0},1})+\sum\limits_{x_{1}\in \{0,1\}}F(\rho _{1,0,x_{1}},\rho
_{1,1,x_{1}}).
\end{equation}%
Combining the Fuchs-van de Graaf inequalities%
\begin{equation}
1-\frac{1}{2}\left\Vert \rho -\xi \right\Vert _{Tr}\leq F(\rho ,\xi )\leq
\sqrt{1-\frac{1}{4}\left\Vert \rho -\xi \right\Vert _{Tr}^{2}}
\end{equation}%
with equations (\ref{qbc88-pb}) and (\ref{qbc88-pa}), the security bound $%
2P_{Alice}^{\ast }+P_{Bob}^{\ast }\geq 2$\ is finally obtained.

This bound was also shown to be optimal, as Ref. \cite{qbc88} exhibited a
family of protocols whose cheating probabilities can be made arbitrarily
close to any point on the $P_{Alice}^{\ast }$ versus $P_{Bob}^{\ast }$\
tradeoff curve.

\section{The protocol}

\subsection{Limitation of the cheating strategy in existing protocols}

Intriguingly, while the security bound $2P_{Alice}^{\ast }+P_{Bob}^{\ast
}\geq 2$\ indicates that in a protocol where Bob cannot cheat (i.e., $%
P_{Bob}^{\ast }=1/2$), Alice can guess Bob's choice $b$ at least with the
probability $P_{Alice}^{\ast }=3/4$, we must note that her cheating strategy
has a serious drawback. That is, once Alice applies the cheating, she will
not be able to determine the values of $x_{0}$, $x_{1}$. For example,
consider the Chailloux-Kerenidis-Sikora (CKS)\ protocol proposed in Sect. 4
of \cite{qbc14} (also presented in Sect. 3.2 of \cite{qbc88} with reverse
usage of the names \textquotedblleft Alice\textquotedblright\ and
\textquotedblleft Bob\textquotedblright ), as described below.

\bigskip

\textit{The CKS protocol}

1. Bob randomly chooses $b\in \{0,1\}$ and prepares the two-qutrit state $%
\left\vert \phi _{b}\right\rangle =(\left\vert bb\right\rangle +\left\vert
22\right\rangle )/\sqrt{2}$. He sends one of the qutrits to Alice.

2. Alice chooses $x_{0},x_{1}\in \{0,1\}$ and applies the unitary
transformation $\left\vert 0\right\rangle \rightarrow (-1)^{x_{0}}\left\vert
0\right\rangle $, $\left\vert 1\right\rangle \rightarrow
(-1)^{x_{1}}\left\vert 1\right\rangle $, $\left\vert 2\right\rangle
\rightarrow \left\vert 2\right\rangle $ on Bob's qutrit.

3. Alice returns the qutrit to Bob who now has the state $\left\vert \psi
_{b}\right\rangle =[(-1)^{x_{b}}\left\vert bb\right\rangle +\left\vert
22\right\rangle ]/\sqrt{2}$.

4. Bob performs the measurement $\{\Pi _{0}=\left\vert \phi
_{b}\right\rangle \left\langle \phi _{b}\right\vert ,\Pi _{1}=\left\vert
\phi _{b}^{\prime }\right\rangle \left\langle \phi _{b}^{\prime }\right\vert
,I-\Pi _{0}-\Pi _{1}\}$\ on the state $\left\vert \psi _{b}\right\rangle $,
where $\left\vert \phi _{b}^{\prime }\right\rangle =(\left\vert
bb\right\rangle -\left\vert 22\right\rangle )/\sqrt{2}$.

5. If the outcome is $\Pi _{0}$ then Bob learns with certainty that $x_{b}=0$%
, if it is $\Pi _{1}$ then $x_{b}=1$, otherwise he aborts.

\bigskip

This protocol can reach $P_{Bob}^{\ast }=1/2$, but it is insecure against
Alice's individual attacks. As shown in Sect. 4 of \cite{qbc14}, Alice's
optimal cheating strategy is simply to measure the qutrit she received in
step 2 using the computational basis. If she gets outcome $\left\vert
0\right\rangle $ ($\left\vert 1\right\rangle $) then she knows with
certainty that $b=0$ ($b=1$). If she gets outcome $\left\vert 2\right\rangle
$ then she guesses the value of $b$. Therefore on average, Alice can learn
Bob's $b$ correctly with the probability $P_{Alice}^{\ast }=3/4$. After the
measurement she returns the measured qutrit to Bob. Then Bob's state will be
either $\left\vert bb\right\rangle $ or $\left\vert 22\right\rangle $. With
any of these two states, the outcome of Bob's measurement in step 4 will
always be either $\Pi _{0}$ or $\Pi _{1}$. Hence he will never abort, so
that Alice's cheating cannot be detected at all.

However, we can see that Alice cannot control, nor she can learn what will
be the actual outcome of Bob's measurement in step 4, because $\left\vert
bb\right\rangle $ and $\left\vert 22\right\rangle $ can both be projected as
$\Pi _{0}$ or $\Pi _{1}$. As a result, at the end of the protocol Bob gets a
bit $x_{b}$, but its value is unknown to Alice. That is, once
dishonest-Alice gains the information on $b$, she loses the information on $%
x_{b}$. Now we prove that this result is general for any Alice's cheating
strategy.

\bigskip

\textbf{Theorem 1}: In the CKS protocol, Alice cannot learn $x_{b}$ with
reliability $1$ and gain a non-trivial amount of information on $b$
simultaneously, while escaping Bob's detection with probability $1$.

\textbf{Proof:} Let $\alpha $ denote the ancillary system that
dishonest-Alice introduced for her cheating, system $\beta $ denote the
qutrit that she received and then returns to Bob, and system $\beta ^{\prime
}$ denote the qutrit that Bob always keeps at his side. To ensure that Bob
will never abort in step 5 so that Alice can pass Bob's detection with
probability $1$, the state of $\beta \otimes \beta ^{\prime }$ at this stage
must be completely contained in the Hilbert space supported by $\left\vert
\phi _{b}\right\rangle _{\beta \beta ^{\prime }}$ and $\left\vert \phi
_{b}^{\prime }\right\rangle _{\beta \beta ^{\prime }}$. Therefore, for any
Alice's cheating strategy, at the end of step 3 the general form of the
quantum system shared by Alice and Bob can always be written as%
\begin{equation}
T(\left\vert e_{ini}\right\rangle _{\alpha }\left\vert \phi
_{b}\right\rangle _{\beta \beta ^{\prime }})=\lambda _{b}^{(0)}\left\vert
e_{b}^{(0)}\right\rangle _{\alpha }\left\vert \phi _{b}\right\rangle _{\beta
\beta ^{\prime }}+\lambda _{b}^{(1)}\left\vert e_{b}^{(1)}\right\rangle
_{\alpha }\left\vert \phi _{b}^{\prime }\right\rangle _{\beta \beta ^{\prime
}},  \label{resultant2}
\end{equation}%
for $b=0,1$, with $\left\vert e_{ini}\right\rangle _{\alpha }$\ ($\left\vert
e_{b}^{(0)}\right\rangle _{\alpha }$ and $\left\vert
e_{b}^{(1)}\right\rangle _{\alpha }$) being the normalized initial (final)
state(s) of $\alpha $, $\left\vert \lambda _{b}^{(0)}\right\vert
^{2}+\left\vert \lambda _{b}^{(1)}\right\vert ^{2}=1$, and $T$\ is the
operator that Alice applies for her cheating. Also, in step (4) of the
protocol, Bob learns that $x_{b}=0$ ($x_{b}=1$) if he gets $\left\vert \phi
_{b}\right\rangle _{\beta \beta ^{\prime }}$\ ($\left\vert \phi _{b}^{\prime
}\right\rangle _{\beta \beta ^{\prime }}$). Therefore, if Alice wants to be
able to learn $x_{b}$ with reliability $1$, she has to choose an operation $%
T $ which can ensure that $\left\vert e_{b}^{(0)}\right\rangle _{\alpha }$
and $\left\vert e_{b}^{(1)}\right\rangle _{\alpha }$\ are orthogonal.

By substituting $\left\vert \phi _{b}\right\rangle _{\beta \beta ^{\prime
}}=(\left\vert bb\right\rangle _{\beta \beta ^{\prime }}+\left\vert
22\right\rangle _{\beta \beta ^{\prime }})/\sqrt{2}$ and $\left\vert \phi
_{b}^{\prime }\right\rangle _{\beta \beta ^{\prime }}=(\left\vert
bb\right\rangle _{\beta \beta ^{\prime }}-\left\vert 22\right\rangle _{\beta
\beta ^{\prime }})/\sqrt{2}$ into equation (\ref{resultant2}), we can
rewrite it as%
\begin{equation}
T(\left\vert e_{ini}\right\rangle _{\alpha }\left\vert \phi
_{b}\right\rangle _{\beta \beta ^{\prime }})=(\left\vert f_{b}\right\rangle
_{\alpha }\left\vert bb\right\rangle _{\beta \beta ^{\prime }}+\left\vert
f_{b}^{\prime }\right\rangle _{\alpha }\left\vert 22\right\rangle _{\beta
\beta ^{\prime }})/\sqrt{2},  \label{resultant}
\end{equation}%
where%
\begin{eqnarray}
\left\vert f_{b}\right\rangle _{\alpha } &\equiv &\lambda
_{b}^{(0)}\left\vert e_{b}^{(0)}\right\rangle _{\alpha }+\lambda
_{b}^{(1)}\left\vert e_{b}^{(1)}\right\rangle _{\alpha },  \nonumber \\
\left\vert f_{b}^{\prime }\right\rangle _{\alpha } &\equiv &\lambda
_{b}^{(0)}\left\vert e_{b}^{(0)}\right\rangle _{\alpha }-\lambda
_{b}^{(1)}\left\vert e_{b}^{(1)}\right\rangle _{\alpha },  \label{fbx}
\end{eqnarray}%
To gain a non-trivial amount of information on $b$, equation (\ref{resultant}%
) indicates that Alice needs to distinguish the reduced density matrices%
\begin{equation}
\rho _{b=0}\equiv (\left\vert f_{0}\right\rangle _{\alpha }\left\langle
f_{0}\right\vert +\left\vert f_{0}^{\prime }\right\rangle _{\alpha
}\left\langle f_{0}^{\prime }\right\vert )/2  \label{rho1}
\end{equation}%
and%
\begin{equation}
\rho _{b=1}\equiv (\left\vert f_{1}\right\rangle _{\alpha }\left\langle
f_{1}\right\vert +\left\vert f_{1}^{\prime }\right\rangle _{\alpha
}\left\langle f_{1}^{\prime }\right\vert )/2.  \label{rho2}
\end{equation}%
While there could exist an operation $T$, which can ensure $\rho _{b=0}\neq
\rho _{b=1}$ before Alice obtains $x_{b}$, we will show below that after
Alice performed any operation $M$ that can make her learn $x_{b}$ with
reliability $1$, $\rho _{b=0}$ and $\rho _{b=1}$ will become equal to each
other, so that they cannot be distinguished any more.

An important fact is that qutrit $\beta ^{\prime }$ is always kept at Bob's
side, so that it remains unchanged under Alice operation $T$. Thus we can
write $T=U_{\alpha \beta }\otimes I_{\beta ^{\prime }}$, where $U_{\alpha
\beta }$ applies on systems $\alpha $ and $\beta $ only, while $I_{\beta
^{\prime }}$\ is the identity operator on $\beta ^{\prime }$. Denoting%
\begin{eqnarray}
U_{\alpha \beta }(\left\vert e_{ini}\right\rangle _{\alpha }\left\vert
0\right\rangle _{\beta }) &=&\left\vert \Psi _{0}\right\rangle _{\alpha
\beta },  \nonumber \\
U_{\alpha \beta }(\left\vert e_{ini}\right\rangle _{\alpha }\left\vert
1\right\rangle _{\beta }) &=&\left\vert \Psi _{1}\right\rangle _{\alpha
\beta },  \nonumber \\
U_{\alpha \beta }(\left\vert e_{ini}\right\rangle _{\alpha }\left\vert
2\right\rangle _{\beta }) &=&\left\vert \Psi _{2}\right\rangle _{\alpha
\beta },  \label{U}
\end{eqnarray}%
then we have
\begin{eqnarray}
T(\left\vert e_{ini}\right\rangle _{\alpha }\left\vert \phi
_{b}\right\rangle _{\beta \beta ^{\prime }}) &=&(U_{\alpha \beta }\otimes
I_{\beta ^{\prime }})(\left\vert e_{ini}\right\rangle _{\alpha }(\left\vert
bb\right\rangle _{\beta \beta ^{\prime }}+\left\vert 22\right\rangle _{\beta
\beta ^{\prime }})/\sqrt{2})  \nonumber \\
&=&(\left\vert \Psi _{b}\right\rangle _{\alpha \beta }\left\vert
b\right\rangle _{\beta ^{\prime }}+\left\vert \Psi _{2}\right\rangle
_{\alpha \beta }\left\vert 2\right\rangle _{\beta ^{\prime }})/\sqrt{2}).
\end{eqnarray}%
Comparing with equation (\ref{resultant}), we yield%
\begin{eqnarray}
\left\vert \Psi _{b}\right\rangle _{\alpha \beta } &=&\left\vert
f_{b}\right\rangle _{\alpha }\left\vert b\right\rangle _{\beta },  \nonumber
\\
\left\vert \Psi _{2}\right\rangle _{\alpha \beta } &=&\left\vert
f_{b}^{\prime }\right\rangle _{\alpha }\left\vert 2\right\rangle _{\beta }.
\end{eqnarray}%
The latter indicates that $\left\vert f_{b}^{\prime }\right\rangle _{\alpha
} $ does not depend on $b$, i.e.,%
\begin{equation}
\left\vert f_{0}^{\prime }\right\rangle _{\alpha }=\left\vert f_{1}^{\prime
}\right\rangle _{\alpha }.  \label{jnt}
\end{equation}%
Then%
\begin{equation}
\left\langle f_{0}^{\prime }\right. \left\vert f_{1}^{\prime }\right\rangle
_{\alpha }=1.  \label{f'f'1}
\end{equation}%
With equation (\ref{fbx}) we know%
\begin{eqnarray}
\left\langle f_{0}^{\prime }\right. \left\vert f_{1}^{\prime }\right\rangle
_{\alpha } &=&(\lambda _{0}^{(0)\ast }\left\langle e_{0}^{(0)}\right\vert
_{\alpha }-\lambda _{0}^{(1)\ast }\left\langle e_{0}^{(1)}\right\vert
_{\alpha })(\lambda _{1}^{(0)}\left\vert e_{1}^{(0)}\right\rangle _{\alpha
}-\lambda _{1}^{(1)}\left\vert e_{1}^{(1)}\right\rangle _{\alpha })
\nonumber \\
&=&\lambda _{0}^{(0)\ast }\lambda _{1}^{(0)}\left\langle e_{0}^{(0)}\right.
\left\vert e_{1}^{(0)}\right\rangle _{\alpha }-\lambda _{0}^{(0)\ast
}\lambda _{1}^{(1)}\left\langle e_{0}^{(0)}\right. \left\vert
e_{1}^{(1)}\right\rangle _{\alpha }  \nonumber \\
&&-\lambda _{0}^{(1)\ast }\lambda _{1}^{(0)}\left\langle e_{0}^{(1)}\right.
\left\vert e_{1}^{(0)}\right\rangle _{\alpha }+\lambda _{0}^{(1)\ast
}\lambda _{1}^{(1)}\left\langle e_{0}^{(1)}\right. \left\vert
e_{1}^{(1)}\right\rangle _{\alpha }  \label{f'f'}
\end{eqnarray}%
and%
\begin{eqnarray}
\left\langle f_{0}\right. \left\vert f_{1}\right\rangle _{\alpha }
&=&(\lambda _{0}^{(0)\ast }\left\langle e_{0}^{(0)}\right\vert _{\alpha
}+\lambda _{0}^{(1)\ast }\left\langle e_{0}^{(1)}\right\vert _{\alpha
})(\lambda _{1}^{(0)}\left\vert e_{1}^{(0)}\right\rangle _{\alpha }+\lambda
_{1}^{(1)}\left\vert e_{1}^{(1)}\right\rangle _{\alpha })  \nonumber \\
&=&\lambda _{0}^{(0)\ast }\lambda _{1}^{(0)}\left\langle e_{0}^{(0)}\right.
\left\vert e_{1}^{(0)}\right\rangle _{\alpha }+\lambda _{0}^{(0)\ast
}\lambda _{1}^{(1)}\left\langle e_{0}^{(0)}\right. \left\vert
e_{1}^{(1)}\right\rangle _{\alpha }  \nonumber \\
&&+\lambda _{0}^{(1)\ast }\lambda _{1}^{(0)}\left\langle e_{0}^{(1)}\right.
\left\vert e_{1}^{(0)}\right\rangle _{\alpha }+\lambda _{0}^{(1)\ast
}\lambda _{1}^{(1)}\left\langle e_{0}^{(1)}\right. \left\vert
e_{1}^{(1)}\right\rangle _{\alpha }.  \label{ff}
\end{eqnarray}%
Now if Alice performs any operation $M$ on the state $T(\left\vert
e_{ini}\right\rangle _{\alpha }\left\vert \phi _{b}\right\rangle _{\beta
\beta ^{\prime }})$\ so that $x_{b}$ ($b=0,1$) is obtained with reliability $%
1$, the final state can still be written as equation (\ref{resultant2}),
except that the coefficients $\lambda _{0}^{(0)}$, $\lambda _{0}^{(1)}$, $%
\lambda _{1}^{(0)}$\ and $\lambda _{1}^{(1)}$\ cannot stay non-vanishing
simultaneously. Instead, one of $\lambda _{0}^{(0)}$ and $\lambda _{0}^{(1)}$
must become zero, and one of $\lambda _{1}^{(0)}$\ and $\lambda _{1}^{(1)}$\
must be zero too. In this case, we can see that in the right hand side of
either equation (\ref{f'f'}) or (\ref{ff}), only one of the coefficients
before the four terms $\left\langle e_{0}^{(0)}\right. \left\vert
e_{1}^{(0)}\right\rangle _{\alpha }$, $\left\langle e_{0}^{(0)}\right.
\left\vert e_{1}^{(1)}\right\rangle _{\alpha }$, $\left\langle
e_{0}^{(1)}\right. \left\vert e_{1}^{(0)}\right\rangle _{\alpha }$ and $%
\left\langle e_{0}^{(1)}\right. \left\vert e_{1}^{(1)}\right\rangle _{\alpha
}$ can remain non-vanishing. No matter which single term remains, there will
always be either%
\begin{equation}
\left\langle f_{0}\right. \left\vert f_{1}\right\rangle _{\alpha
}=\left\langle f_{0}^{\prime }\right. \left\vert f_{1}^{\prime
}\right\rangle _{\alpha }.
\end{equation}%
or%
\begin{equation}
\left\langle f_{0}\right. \left\vert f_{1}\right\rangle _{\alpha
}=-\left\langle f_{0}^{\prime }\right. \left\vert f_{1}^{\prime
}\right\rangle _{\alpha }.
\end{equation}%
Combining with equation (\ref{f'f'1}), they both give%
\begin{equation}
\left\vert f_{0}\right\rangle _{\alpha }\left\langle f_{0}\right\vert
=\left\vert f_{1}\right\rangle _{\alpha }\left\langle f_{1}\right\vert .
\end{equation}%
Substituting it and equation (\ref{jnt}) into equations (\ref{rho1}) and (%
\ref{rho2}), we finally obtain%
\begin{equation}
\rho _{b=0}=\rho _{b=1}.
\end{equation}%
Thus they provide absolutely zero knowledge on $b$. Therefore, once Alice
performs the operation to learn $x_{b}$ with reliability $1$, she can no
longer gain any information on $b$. This ends the proof of Theorem 1.

\bigskip

The above proof does not exclude the existence of other cheating strategies,
in which Alice can learn $x_{b}$ with a reliability less than $1$, and/or
Bob may abort in step 5 with a non-vanishing probability. But this will do
no harm to our purpose, as it will be shown later in Sects. IV.B and IV.C.

In fact, besides the CKS protocol, some other QOT protocols \cite%
{HeQOT,qi135} also display the same feature described in Theorem 1. Let $%
(3/4,1/2)$\textit{-protocol} denote any QOT protocol of this kind, i.e.,
both $P_{Alice}^{\ast }=3/4$ and $P_{Bob}^{\ast }=1/2$\ are satisfied
exactly, and Alice cannot determine $x_{b}$ with reliability $1$ once she
gain a non-trivial amount of information on $b$. It will be shown below that
though a $(3/4,1/2)$-protocol merely saturates the security bound $%
2P_{Alice}^{\ast }+P_{Bob}^{\ast }\geq 2$, it can be utilized to construct a
compound protocol which can eventually violate this bound when Alice is
limited to individual attacks.

\subsection{Our protocol}

Our method is to use such a $(3/4,1/2)$-protocol as a building block, with
which Alice transfers a series of bits $x_{0}^{(i)}$, $x_{1}^{(i)}$ (not the
final $x_{0}$, $x_{1}$ that she wants to transfer) to Bob. The values of $%
x_{0}^{(i)}$, $x_{1}^{(i)}$ are \textit{not} completely random. Instead,
they must be chosen according to a certain rule. Then Bob uses many of $%
x_{0}^{(i)}$, $x_{1}^{(i)}$ to check whether Alice can determine their
values correctly. Finally he uses one of the remaining pairs of $x_{0}^{(i)}$%
, $x_{1}^{(i)}$ and asks Alice to encode her $x_{0}$, $x_{1}$. The general
form of the protocol is as follows.

\bigskip

\textit{Protocol A: weak OT for transferring }$(x_{0},x_{1})$

A1. Alice and Bob discuss and agree on a set $S$ of classical $n$-bit
strings.

A2. Alice randomly chooses two strings\newline
$X_{0}=x_{0}^{(1)}x_{0}^{(2)}...x_{0}^{(i)}...x_{0}^{(n)}$ and $%
X_{1}=x_{1}^{(1)}x_{1}^{(2)}...x_{1}^{(i)}...x_{1}^{(n)}$ from $S$. Note
that at this stage, none of these $x_{0}^{(i)}$, $x_{1}^{(i)}$ have any
specific relationship with the two final bits $x_{0}$, $x_{1}$ (we call them
as \textit{target bits} thereafter) that Alice wants to transfer to Bob as
the goal of the weak OT.

A3. For each $i$ ($i=1,...,n$), Alice transfers $x_{0}^{(i)}$, $x_{1}^{(i)}$
to Bob using a $(3/4,1/2)$-protocol. Bob randomly chooses $b_{i}\in \{0,1\}$
and decodes $x_{b_{i}}$.

A4. Security check: among all these $n$ runs of the 
$(3/4,1/2)$-protocol, Bob picks $m$ ($m<n$) runs randomly. For each of these
runs, he asks Alice to announce $x_{0}^{(i)}$ and $x_{1}^{(i)}$, and checks
whether they are consistent with the value of $x_{b_{i}}$ that he obtained
in the $(3/4,1/2)$-protocol. He also checks that there is at least two
strings $X_{0}^{\prime } $ and $X_{1}^{\prime }$ in set $S$, such that all
the $m$ bits $x_{0}^{(i)} $ ($x_{1}^{(i)}$) that Alice announced are
contained in $X_{0}^{\prime }$ ($X_{1}^{\prime }$).

A5. If Alice's announced values pass the above check, Bob picks one of the
remaining $n-m$ unchecked runs (which is denoted as the $\hat{i}$-th run) of
the $(3/4,1/2)$-protocol. This run should satisfy the requirement that $%
x_{0}^{(\hat{i})}=0$, $x_{0}^{(\hat{i})}=1$, $x_{1}^{(\hat{i})}=0$, and $%
x_{1}^{(\hat{i})}=1$ are all allowed by set $S$. That is, in set $S$ there
is at least one string which contains all the $m$ bits $x_{0}^{(i)}$ ($%
x_{1}^{(i)}$) that Alice announced in step A4 and $x_{0}^{(\hat{i})}=0$ ($%
x_{1}^{(\hat{i})}=0$), and an equal number of strings, each of which also
contains all these $m$ bits $x_{0}^{(i)}$ ($x_{1}^{(i)}$) but with $x_{0}^{(%
\hat{i})}=1$ ($x_{1}^{(\hat{i})}=1$) instead. This guarantees that the $m$
bits $x_{0}^{(i)}$ ($x_{1}^{(i)}$) announced in step A4 are insufficient for
Bob to deduce the value of $x_{0}^{(\hat{i})}$ ($x_{1}^{(\hat{i})}$) from
set $S$. Alice checks that this requirement is met after Bob told her the
value of $\hat{i}$.

A6. Alice completes the weak OT by using the $\hat{i}$-th run of the $%
(3/4,1/2)$-protocol\ to encode the target bits $x_{0}$, $x_{1}$. That is,
she announces $x_{0}\oplus x_{0}^{(\hat{i})}$ and $x_{1}\oplus x_{1}^{(\hat{i%
})}$ to Bob. Thus Bob can obtain either $x_{0}$ or $x_{1}$ depending on
whether he has obtained $x_{0}^{(\hat{i})}$ or $x_{1}^{(\hat{i})}$ in the $%
\hat{i}$-th run of the $(3/4,1/2)$-protocol.

\bigskip

\subsection{A concrete example}

To make our protocol easier for understanding and analyzing, here we provide
a concrete example of our above protocol where the CKS protocol is used as
the $(3/4,1/2)$-protocol and the explicit form of set $S$ is given.

\bigskip

\textit{Protocol B: a concrete example}

B1. Alice and Bob run the CKS protocol for $n=3k$\ times. Every $3$ runs of
the CKS protocol are grouped together and we call it as a \textit{triple run}%
. Let $x_{0}^{(i)}$, $x_{1}^{(i)}$ ($i=1,2,3$) denote the bits that Alice
transfers to Bob in a triple run. The values of the strings $%
X_{0}=x_{0}^{(1)}x_{0}^{(2)}x_{0}^{(3)}$ and $%
X_{1}=x_{1}^{(1)}x_{1}^{(2)}x_{1}^{(3)}$ cannot be completely random.
Instead, they are required to be chosen within the set $S=\{000,001,010,100%
\} $.

B2. Security check: for every triple run, Bob randomly picks two runs of the
CKS protocol, denotes them as the $i_{1}$-th and $i_{2}$-th runs. The
remaining run that is not picked is denoted as the $i_{3}$-th run. Bob asks
Alice to reveal $x_{0}^{(i_{1})}$, $x_{1}^{(i_{1})}$ and $x_{0}^{(i_{2})}$, $%
x_{1}^{(i_{2})}$. If $%
x_{0}^{(i_{1})}=x_{1}^{(i_{1})}=x_{0}^{(i_{2})}=x_{1}^{(i_{2})}=0$\ then Bob
marks the corresponding triple run as a \textit{useful run}, as both $%
x_{0}^{(i_{3})}$ and $x_{1}^{(i_{3})}$\ can either be $0$ or $1$\ according
to the definition of set $S$, so that they may potentially be used for
encoding the target bits $x_{0}$, $x_{1}$ later. Else if any of $%
x_{0}^{(i_{1})}$, $x_{1}^{(i_{1})}$, $x_{0}^{(i_{2})}$, $x_{1}^{(i_{2})}$\
is $1$, Bob asks Alice to reveal $x_{0}^{(i_{3})}$, $x_{1}^{(i_{3})}$ too,
and checks whether both $X_{0}=x_{0}^{(1)}x_{0}^{(2)}x_{0}^{(3)}$ and $%
X_{1}=x_{1}^{(1)}x_{1}^{(2)}x_{1}^{(3)}$ belong to set $S$. He also checks
that none of Alice's announced values conflicts with what he decoded from
the CKS protocol.

B3. If Alice's data passes the above check, Bob picks one of the useful runs
and asks Alice to complete the weak OT using this run. Then Alice announces $%
x_{0}\oplus x_{0}^{(i_{3})}$ and $x_{1}\oplus x_{1}^{(i_{3})}$ to Bob, so
that he can obtain either the target bit $x_{0}$ or $x_{1}$ depending on
whether he has obtained $x_{0}^{(i_{3})}$ or $x_{1}^{(i_{3})}$ in the
corresponding run of the CKS protocol.

B4. For better security, Bob can further ask Alice to reveal $%
x_{0}^{(i_{3})} $ and $x_{1}^{(i_{3})}$\ of all the rest useful runs which
are not picked in step B3. Then he checks whether they conflict with what he
decoded from the CKS protocol.

\bigskip

\section{Security}

\subsection{The collective attack}

The above protocols A and B are, unfortunately, still restricted by the
security bound $2P_{Alice}^{\ast }+P_{Bob}^{\ast }\geq 2$\ if Alice has
unlimited computational power to apply collective attacks. Taking Protocol B
as an example, her cheating strategy is as follows.

In step B1, for each triple run Alice introduces a $6$-qubit system $%
C=c_{0}^{(1)}c_{0}^{(2)}c_{0}^{(3)}c_{1}^{(1)}c_{1}^{(2)}c_{1}^{(3)}$ to
keep her choice of the strings $X_{0}=x_{0}^{(1)}x_{0}^{(2)}x_{0}^{(3)}$ and
$X_{1}=x_{1}^{(1)}x_{1}^{(2)}x_{1}^{(3)}$ at the quantum level. The state of
system $C$ is initialized as%
\begin{eqnarray}
&&\left\vert
c_{0}^{(1)}c_{0}^{(2)}c_{0}^{(3)}c_{1}^{(1)}c_{1}^{(2)}c_{1}^{(3)}\right%
\rangle  \nonumber \\
&=&\frac{1}{2}(\left\vert 000\right\rangle +\left\vert 001\right\rangle
+\left\vert 010\right\rangle +\left\vert 100\right\rangle )  \nonumber \\
&&\otimes \frac{1}{2}(\left\vert 000\right\rangle +\left\vert
001\right\rangle +\left\vert 010\right\rangle +\left\vert 100\right\rangle ),
\label{C}
\end{eqnarray}%
where the first (last) three qubits are corresponding to the string $X_{0}$ (%
$X_{1}$). That is, it is a superposition of all the states allowed by set $S$%
.

In the $i$-th run ($i=1,2,3$) of the CKS protocol during a triple run, let $%
\beta ^{(i)}$\ denote the qutrit that Bob sent to Alice, taken from his
two-qutrit state $\left\vert \phi _{b}^{(i)}\right\rangle =(\left\vert
bb\right\rangle +\left\vert 22\right\rangle )/\sqrt{2}$. Alice uses $%
c_{0}^{(i)}$, $c_{1}^{(i)}$ as control qubits to determine her
transformation on $\beta ^{(i)}$. That is, on $c_{0}^{(i)}\otimes
c_{1}^{(i)}\otimes \beta ^{(i)}$ she applies the unitary transformation%
\begin{eqnarray}
T_{c_{0}^{(i)}c_{1}^{(i)}\beta ^{(i)}}
&=&\sum\limits_{x_{0}^{(i)},x_{1}^{(i)}=0}^{1}(\left\vert
x_{0}^{(i)}\right\rangle _{c_{0}^{(i)}}\left\langle x_{0}^{(i)}\right\vert
\otimes \left\vert x_{1}^{(i)}\right\rangle _{c_{1}^{(i)}}\left\langle
x_{1}^{(i)}\right\vert  \nonumber \\
&&\otimes \left[
\begin{array}{ccc}
(-1)^{x_{0}^{(i)}} & 0 & 0 \\
0 & (-1)^{x_{1}^{(i)}} & 0 \\
0 & 0 & 1%
\end{array}%
\right] _{\beta ^{(i)}}).  \label{T}
\end{eqnarray}%
By doing so, Alice manages to finish the transformation $\left\vert
0\right\rangle \rightarrow (-1)^{x_{0}}\left\vert 0\right\rangle $, $%
\left\vert 1\right\rangle \rightarrow (-1)^{x_{1}}\left\vert 1\right\rangle $%
, $\left\vert 2\right\rangle \rightarrow \left\vert 2\right\rangle $ on
Bob's qutrit $\beta ^{(i)}$, just as it is required in the CKS protocol when
Alice is honest. The only difference is that in the current case, $%
x_{0}^{(i)}$\ and $x_{1}^{(i)}$\ do not have deterministic classical values.
Instead, they are kept at the quantum level.

In step B2 whenever Bob picks one run of the CKS protocol and asks Alice to
reveal the corresponding $x_{0}^{(i)}$\ and $x_{1}^{(i)}$, Alice measures
the qubits $c_{0}^{(i)}$ and $c_{1}^{(i)}$ in the computational basis $%
\{\left\vert 0\right\rangle ,\left\vert 1\right\rangle \}$. If the result is
$\left\vert 0\right\rangle $ ($\left\vert 1\right\rangle $) then she
announces the corresponding $x_{0}^{(i)}$\ or $x_{1}^{(i)}$ as $0$ ($1$).
From equation (\ref{T}) it can be seen that Alice's announcement will never
conflict with the values Bob decodes from the CKS protocol. Now recall that
a useful run is defined as the triple run where $%
x_{0}^{(i_{1})}=x_{1}^{(i_{1})}=x_{0}^{(i_{2})}=x_{1}^{(i_{2})}=0$.
Therefore by combining equations (\ref{C}) and (\ref{T}), we know that the
state of $c_{0}^{(i_{3})}\otimes c_{1}^{(i_{3})}\otimes \phi _{b}^{(i_{3})}$
of any useful run at the end of step B2 becomes%
\begin{eqnarray}
&&\left\vert c_{0}^{(i_{3})}\otimes c_{1}^{(i_{3})}\otimes \phi
_{b}^{(i_{3})}\right\rangle  \nonumber \\
&=&\frac{1}{2}\sum\limits_{x_{0}^{(i_{3})},x_{1}^{(i_{3})}=0}^{1}\{\left%
\vert x_{0}^{(i_{3})}\right\rangle _{c_{0}^{(i_{3})}}\otimes \left\vert
x_{1}^{(i_{3})}\right\rangle _{c_{1}^{(i_{3})}}  \nonumber \\
&&\otimes \frac{1}{\sqrt{2}}[(-1)^{x_{b}^{(i_{3})}}\left\vert
bb\right\rangle +\left\vert 22\right\rangle ]\}.  \label{B2}
\end{eqnarray}

If a useful run is picked for the security check in step B4, Alice can
simply measure the qubits $c_{0}^{(i_{3})}$ and $c_{1}^{(i_{3})}$ in the
basis $\{\left\vert 0\right\rangle ,\left\vert 1\right\rangle \}$ and reveal
$x_{0}^{(i_{3})}$ and $x_{1}^{(i_{3})}$\ correctly. On the other hand, if a
useful run is picked in step B3 to encode Alice's target bits $x_{0}$, $%
x_{1} $, then she will have the freedom to choose whether to measure $%
c_{0}^{(i_{3})}$ and $c_{1}^{(i_{3})}$ in the basis $\{\left\vert
0\right\rangle ,\left\vert 1\right\rangle \}$ and learn the values of $x_{0}$%
, $x_{1}$\ as an honest Alice does, or to learn the value of Bob's $b$ with
a certain probability instead. In the latter case, she measures $%
c_{0}^{(i_{3})}$ and $c_{1}^{(i_{3})}$ in the basis $\{\left\vert
+\right\rangle ,\left\vert -\right\rangle \}$, where $\left\vert \pm
\right\rangle =(\left\vert 0\right\rangle \pm \left\vert 1\right\rangle )/%
\sqrt{2}$. This is because equation (\ref{B2}) can be rewritten as%
\begin{eqnarray}
&&\left\vert c_{\bar{b}}^{(i_{3})}\otimes c_{b}^{(i_{3})}\otimes \phi
_{b}^{(i_{3})}\right\rangle  \nonumber \\
&=&\frac{1}{2}\sum\limits_{x_{b}^{(i_{3})}=0}^{1}\{(\sum\limits_{x_{\bar{b}%
}^{(i_{3})}=0}^{1}\left\vert x_{\bar{b}}^{(i_{3})}\right\rangle _{c_{\bar{b}%
}^{(i_{3})}})\otimes \left\vert x_{b}^{(i_{3})}\right\rangle
_{c_{b}^{(i_{3})}}  \nonumber \\
&&\otimes \frac{1}{\sqrt{2}}[(-1)^{x_{b}^{(i_{3})}}\left\vert
bb\right\rangle +\left\vert 22\right\rangle ]\}  \nonumber \\
&=&\frac{1}{\sqrt{2}}\left\vert +\right\rangle _{c_{\bar{b}%
}^{(i_{3})}}\otimes \sum\limits_{x_{b}^{(i_{3})}=0}^{1}\{\left\vert
x_{b}^{(i_{3})}\right\rangle _{c_{b}^{(i_{3})}}  \nonumber \\
&&\otimes \frac{1}{\sqrt{2}}[(-1)^{x_{b}^{(i_{3})}}\left\vert
bb\right\rangle +\left\vert 22\right\rangle ]\}  \nonumber \\
&=&\frac{1}{\sqrt{2}}\left\vert +\right\rangle _{c_{\bar{b}%
}^{(i_{3})}}\otimes \lbrack \left\vert -\right\rangle
_{c_{b}^{(i_{3})}}\otimes \left\vert bb\right\rangle +\left\vert
+\right\rangle _{c_{b}^{(i_{3})}}\otimes \left\vert 22\right\rangle ].
\end{eqnarray}%
We can see that if Alice finds the outcome of her measurement on $%
c_{0}^{(i_{3})}$ ($c_{1}^{(i_{3})}$) is $\left\vert -\right\rangle $, then
she knows with certainty that Bob's choice is $b=0$\ ($b=1$). This will
occur with the probability $1/2$. On the other $1/2$ case, the outcomes of
Alice's measurements on both $c_{0}^{(i_{3})}$ and $c_{1}^{(i_{3})}$ are $%
\left\vert +\right\rangle $, thus she has to guess the value of $b$ by
herself. Therefore, the average probability that Alice can learn Bob's $b$
correctly is still $P_{Alice}^{\ast }=3/4$, which is the same as that of the
original CKS protocol. As there is always $P_{Bob}^{\ast }\geq 1/2$ for any
protocol, we can see that the security bound $2P_{Alice}^{\ast
}+P_{Bob}^{\ast }\geq 2$ still holds in the current case.

\subsection{Security against individual attacks}

However, the above cheating requires the computational power to perform
collective operations on many qubits/qutrits. More rigorously, equations (%
\ref{C}) and (\ref{T}) indicate that at the end of step B1, in every triple
run Alice needs to make $6$ qubits and $3$ qutrits entangled together, even
if Bob's half of his two-qutrit states is not counted. Here we will show
that if Alice is limited to individual measurements, then the protocol can
be secure.

In this scenario, during each run of the CKS protocol, Alice is not allowed
to perform collective operations to entangle the qutrit $\beta $ she
received from Bob with her quantum ancillary system anymore. What she can do
is to handle $\beta $ individually. In general, such operations can be
modeled as a channel $C_{\alpha \beta }$ which takes $\beta $ as an input,
then outputs a single qutrit state and a classical register $\alpha $
containing her measurement outcome. The effect of the channel $C_{\alpha
\beta }$ can be written as%
\begin{equation}
C_{\alpha \beta }(\left\vert e_{ini}\right\rangle _{\alpha }\left\vert
j\right\rangle _{\beta })=\sum\limits_{j^{\prime }=0}^{2}\lambda
_{jj^{\prime }}\left\vert e_{jj^{\prime }}\right\rangle _{\alpha }\left\vert
j^{\prime }\right\rangle _{\beta },
\end{equation}%
where $j=0,1,2$, and $\left\vert e_{ini}\right\rangle _{\alpha }$\ ($%
\left\vert e_{jj^{\prime }}\right\rangle _{\alpha }$) is the initial (final)
state of $\alpha $. Comparing with equations (\ref{resultant2}) and (\ref{U}%
), we can see that $C_{\alpha \beta }$ is actually a special case of the
general cheating operation $T$ studied in the proof of Theorem 1. The
specialty in the current case is that $\alpha $ is classical, while in
equation (\ref{resultant2}) it can be either classical or quantum.
Therefore, by formulating the resultant state of $C_{\alpha \beta }$ as
equation (\ref{resultant2}) and repeating the same proof in Sect. III.A, we
find that the result of Theorem 1 still applies here. That is, if Alice can
guess $b$ with nonzero bias by applying channel $C_{\alpha \beta }$ to Bob's
qutrit $\beta $ and Bob never aborts, then she cannot learn $x_{b}$ with
reliability $1$.

Moreover, as $\alpha $ is a classical register, there will be no alternative
bases for measuring it. That is, once Alice decides on what kind of channel
to apply, then the measurement basis for $\alpha $ is also fixed. No matter
when Alice will measure $\alpha $ and extract the information stored in it,
this information is already a deterministic classical object after the
channel is applied, and there is only one choice of the basis for extracting
it. This is different from an unlimited quantum attack, where Alice can
apply the cheating operation and delay the measurement, then at a later
time, if she wants to learn $x_{b}$, she measures $\alpha $ in a certain
basis, while if she wants to learn $b$, she measures $\alpha $ in another
basis. In the current case, even if the measurement could be delayed, there
is only one basis for Alice (otherwise it will become a collective attack).
Therefore, Alice needs to determine beforehand which basis to use, and picks
the corresponding channel to apply. Dishonest-Alice will surely choose a
basis which enables her to learn Bob's $b$ with a nonzero bias, because this
is the goal of her cheating. But then Theorem 1 guarantees that she cannot
know\ with certainty the value of $x_{b}$ that Bob actually obtained.
Consequently, if this run of the CKS protocol is picked for the security
check in Protocol B, Alice will stand a non-vanishing probability $%
\varepsilon $ to either announce a wrong value of $x_{b}$ or cause Bob to
abort (in case his measurement outcome is neither $\Pi _{0}$ nor $\Pi _{1}$
in step 5 of the CKS protocol).

Now suppose that Alice chooses to cheat in $pn$ ($1/n\leq p\leq 1$) runs of
the CKS protocol. While Alice can apply different strategies in these runs
so that the value of $\varepsilon $ can vary, we can define $\varepsilon
_{m} $\ as the minimum of $\varepsilon $ in any run that Alice cheats. Thus $%
1-\varepsilon _{m}$\ is the maximal probability for Alice to pass Bob's
check in a single run. Since at the end of Protocol B, $n-1$ runs of the CKS
protocol will be checked, there can be two possibilities.

(I) The only one run that is not checked is picked among the $pn$ runs that
Alice cheats. Since this run is used for encoding the target bits in step
B3, Alice can gain a non-trivial amount of information on $b$. As the CKS
protocol ensures that Alice can learn $b$ correctly with the probability $%
3/4 $ at the most, and the other $pn-1$ runs that Alice cheats are all
checked, the maximal probability for Alice to learn $b$ correctly and pass
the checks successfully in this case is\
\begin{equation}
P_{Alice}^{I}\leq \frac{3}{4}(1-\varepsilon _{m})^{pn-1}.
\end{equation}

(II) The only one run that is not checked is not picked among the $pn$ runs
that Alice cheats. As Alice acts honestly in this run, she can only get $b$
by guess, which can be correct with the probability $1/2$. Meanwhile, all
the $pn$ runs that Alice cheats are checked. Thus the maximal probability
for Alice to learn $b$ correctly and pass the checks successfully in this
case is\
\begin{equation}
P_{Alice}^{II}\leq \frac{1}{2}(1-\varepsilon _{m})^{pn}.
\end{equation}

Note that cases (I) and (II) occur with the probabilities $p$ and $1-p$,
respectively. Thus\ the total probability for Alice to pass the check while
learning $b$ correctly is\
\begin{eqnarray}
P_{Alice}^{\ast } &=&pP_{Alice}^{I}+(1-p)P_{Alice}^{II}  \nonumber \\
&\leq &\frac{3}{4}p(1-\varepsilon _{m})^{pn-1}+\frac{1}{2}%
(1-p)(1-\varepsilon _{m})^{pn}.
\end{eqnarray}%
Since%
\begin{eqnarray}
&&\frac{\partial }{\partial p}(\frac{3}{4}p(1-\varepsilon _{m})^{pn-1}+\frac{%
1}{2}(1-p)(1-\varepsilon _{m})^{pn})  \nonumber \\
&=&((\frac{3}{4}p+\frac{1}{2}(1-p)(1-\varepsilon _{m}))n\ln (1-\varepsilon
_{m})  \nonumber \\
&&+\frac{3}{4}-\frac{1}{2}(1-\varepsilon _{m}))(1-\varepsilon _{m})^{pn-1}
\nonumber \\
&<&0,
\end{eqnarray}%
higher $P_{Alice}^{\ast }$ can be obtained by lowering $p$. The lowest
nonzero $p$ is $p=1/n$, i.e., Alice cheats in $pn=1$ run only and hopes that
she is so lucky that this run is finally picked for encoding the target
bits. In this case\
\begin{eqnarray}
P_{Alice}^{\ast } &\leq &\frac{1}{2}+\frac{1}{4n}-\frac{\varepsilon _{m}}{2}%
(1-\frac{1}{n})  \nonumber \\
&\leq &\frac{1}{2}+\frac{1}{4n}.
\end{eqnarray}%
As a result, for any arbitrarily small positive constant $\zeta $, Bob can
choose $n>1/(4\zeta )$ and ask Alice to perform the corresponding Protocol
B, which can achieve $P_{Alice}^{\ast }<1/2+\zeta $.

On the other hand, Bob's cheating probability remains the same as that of
the CKS protocol. This is because in any useful run, the values of $%
x_{0}^{(i_{1})}$, $x_{1}^{(i_{1})}$, $x_{0}^{(i_{2})}$, $x_{1}^{(i_{2})}$
that Alice revealed are always $0$. As set $S$ is defined as $%
S=\{000,001,010,100\}$, any value of $x_{0}^{(i_{3})}$ and $x_{1}^{(i_{3})}$%
\ remains possible to Bob unless Alice reveals them. Thus the values of $%
x_{0}^{(i_{1})}$, $x_{1}^{(i_{1})}$, $x_{0}^{(i_{2})}$, $x_{1}^{(i_{2})}$ in
a useful run do not provide any information for Bob to deduce $%
x_{0}^{(i_{3})}$ and $x_{1}^{(i_{3})}$. Also, the values of $x_{0}^{(i)}$, $%
x_{1}^{(i)}$ in different triple runs are chosen independently, so that the
specific $x_{0}^{(i_{3})}$, $x_{1}^{(i_{3})}$\ finally chosen for encoding
the target bits $x_{0}$, $x_{1}$ are not affected by any $x_{0}^{(i)}$, $%
x_{1}^{(i)}$ from all the other runs. Consequently, Bob still has to decode
the target bits via the corresponding run of the CKS protocol, without any
help from other runs. Therefore, his cheating probability in our Protocol B
is still $P_{Bob}^{\ast }=1/2$, as what can be obtained in a single run of
the original CKS protocol \cite{qbc14}.

Putting things together, we can see that when Alice is limited to individual
measurements, in our Protocol B $2P_{Alice}^{\ast }+P_{Bob}^{\ast }$ can be
made arbitrarily close to $3/2$, which is the maximal violation of the
security bound $2P_{Alice}^{\ast }+P_{Bob}^{\ast }\geq 2$ since the minimums
for $P_{Alice}^{\ast }$ and $P_{Bob}^{\ast }$\ are both $1/2$.

\subsection{Security against limited collective attacks}

If Alice is allowed to perform collective operations but it is restricted to
a limited number of quantum systems only, then the security bound $%
2P_{Alice}^{\ast }+P_{Bob}^{\ast }\geq 2$ can also be violated to a certain
degree.

Here we consider the case where Alice's collective operations are limited to
the quantum systems in the same run of the CKS protocol only, i.e., the
qutrit $\beta ^{(i)}$ that Bob sends to her and the ancillary system she
introduces (e.g., it can contain the two control qubits $c_{0}^{(i)}$, $%
c_{1}^{(i)}$ for keeping $x_{0}^{(i)}$, $x_{1}^{(i)}$ at the quantum level).
She can still apply the transformation defined in equation (\ref{T}) or
other operations on these systems for cheating. Our discussion below will
remain valid as long as this ancillary system cannot be entangled with the
ancillary system for any other run (e.g., equation (\ref{C}) is not allowed).


In this scenario, since the potential cheating strategies could be
innumerous and much more complicated than the individual attacks, it is hard
to prove the exact security bound of our protocol. But at least here we can
obtain the loose upper and lower bounds of the security, which is $5/3\leq
2P_{Alice}^{\ast }+P_{Bob}^{\ast }<2$. It means that the probability for
successful cheatings is higher than that of the individual attacks, but it
still violates the security bound for the unlimited collective attack.

Let us prove the upper bound $2P_{Alice}^{\ast }+P_{Bob}^{\ast }<2$\ first.
After the end of step B1, from Alice's point of view, the general form of
the state of Alice's and Bob's combined system for each single run of the
CKS protocol is%
\begin{eqnarray}
\left\vert \alpha \otimes \beta \otimes \beta ^{\prime }\right\rangle
&=&\lambda _{b}^{(0)}\left\vert e_{b}^{(0)}\right\rangle _{\alpha
}\left\vert \phi _{b}\right\rangle _{\beta \beta ^{\prime }}+\lambda
_{b}^{(1)}\left\vert e_{b}^{(1)}\right\rangle _{\alpha }\left\vert \phi
_{b}^{\prime }\right\rangle _{\beta \beta ^{\prime }}  \nonumber \\
&&+\lambda _{b}^{(2)}\left\vert e_{b}^{(2)}\otimes \phi _{b}^{\prime \prime
}\right\rangle _{\alpha \beta \beta ^{\prime }},
\end{eqnarray}%
where the notations are the same as those in the proof of Theorem 1, with
the additional $\left\vert e_{b}^{(2)}\otimes \phi _{b}^{\prime \prime
}\right\rangle _{\alpha \beta \beta ^{\prime }}$, which represents the state
orthogonal to both $\left\vert e_{b}^{(0)}\right\rangle _{\alpha }\left\vert
\phi _{b}\right\rangle _{\beta \beta ^{\prime }}$ and $\left\vert
e_{b}^{(1)}\right\rangle _{\alpha }\left\vert \phi _{b}^{\prime
}\right\rangle _{\beta \beta ^{\prime }}$. Note that the actual system may
already collapse to one of the terms at the right hand side of the equation
due to Bob's measurement on $\beta \otimes \beta ^{\prime }$. But Alice can
still treat the whole state as the entangled form in this equation if she
has not measured $\alpha $. This is because Alice's and Bob's local
operations are commutable for the entangled system $\alpha \otimes \beta
\otimes \beta ^{\prime }$, so that it does not matter mathematically who
performs the measurement first.

Since Bob learns that $x_{b}^{(i)}=0$ ($x_{b}^{(i)}=1$) if he gets $%
\left\vert \phi _{b}\right\rangle _{\beta \beta ^{\prime }}$\ ($\left\vert
\phi _{b}^{\prime }\right\rangle _{\beta \beta ^{\prime }}$), otherwise he
aborts, the above equation can be understood as%
\begin{eqnarray}
\left\vert \alpha \otimes \beta \otimes \beta ^{\prime }\right\rangle
&=&\lambda _{b}^{(0)}\left\vert e_{b}^{(0)}\right\rangle _{\alpha
}\left\vert x_{b}^{(i)}=0\right\rangle _{\beta \beta ^{\prime }}+\lambda
_{b}^{(1)}\left\vert e_{b}^{(1)}\right\rangle _{\alpha }\left\vert
x_{b}^{(i)}=1\right\rangle _{\beta \beta ^{\prime }}  \nonumber \\
&&+\lambda _{b}^{(2)}\left\vert e_{b}^{(2)}\otimes abort \right\rangle
_{\alpha \beta \beta ^{\prime }},
\end{eqnarray}%
Comparing with equation (\ref{resultant2}), it is even more general since it
also includes the case where Bob may abort. Now if none of the coefficients $%
\lambda _{b}^{(0)}$\ and $\lambda _{b}^{(1)}$\ equals exactly to $1$, then
the value of $x_{b}^{(i)}$ is kept at the quantum level. That is, it will be
determined by the uncertainty in quantum measurement, so that Alice cannot
control with certainty which value can be\ obtained by Bob. Else if one of $%
\lambda _{b}^{(0)}$\ and $\lambda _{b}^{(1)}$\ equals exactly to $1$, then
the other one and $\lambda _{b}^{(2)}$\ obviously have to be zero, and the
value of $x_{b}^{(i)}$ becomes classically deterministic.

After the end of step B1, suppose that the values of $x_{0}^{(i)}$, $%
x_{1}^{(i)}$ in $pn$ ($n=3k$, $0\leq p\leq 1$) runs of the CKS protocol in
Protocol B are kept at the quantum level (the values of $\lambda _{b}^{(0)}$%
\ and $\lambda _{b}^{(1)}$\ depend on Alice's specific strategy, which can
be different in each run). In the rest $(1-p)n$ runs, $x_{0}^{(i)}$, $%
x_{1}^{(i)}$ are no longer kept at the quantum level after step B1, but take
deterministic classical values instead, so that Alice can ensure that the
values of both $X_{0}$ and $X_{1}$ are presented in set $S$. Then these $%
(1-p)n$ runs are in fact executed honestly, as Theorem 1 ensures that Alice
cannot use them to decode Bob's $b$. She can get $b$ only if one of the
other $pn$ runs of the CKS protocol dishonestly executed is picked in step
B3 for encoding the target bits $x_{0}$ and $x_{1}$ to complete the weak OT.
This will occur with the probability $p$. Even if Alice uses the optimal
cheating strategy so that she can still learn Bob's $b$ with the probability
$3/4$ (which is the maximum that can be obtained in the CKS protocol) for
such a single dishonest run, the probability for (this run to be chosen)
\textit{and} ($b$ is learned correctly) will drop down to $(3/4)p$. If any
other non-optimal cheating strategy was used in this run, the probability is
limited by this value too. In the rest $(1-p)$ occasions where one of the $%
(1-p)n$ honestly executed runs is chosen for encoding the target bits, Alice
can only get Bob's $b$ by guessing which has probability $1/2$ to be
correct. Therefore, the total probability for Alice to cheat successfully in
our Protocol B is bounded by%
\begin{eqnarray}
P_{Alice}^{\ast } &=&[\frac{3}{4}p+\frac{1}{2}(1-p)]p_{c}  \nonumber \\
&=&(\frac{1}{2}+\frac{1}{4}p)p_{c},  \label{pa}
\end{eqnarray}%
where $p_{c}$\ is the probability that Alice can pass the security checks.
Finding the tight bound for $P_{Alice}^{\ast }$\ requires a rigorous
evaluation of $p_{c}$, which will depend on the specific cheating strategy
Alice applies on the $pn$ runs. But we can show that there will always be $%
P_{Alice}^{\ast }<3/4$. This is because if $p<1$, for a loose evaluation we
can simply take the maximum $p_{c}=1$ which surely covers any strategy. Then
$P_{Alice}^{\ast }=1/2+p/4<3/4$. On the other hand, consider the case $p=1$.
As the collective attacks are limited to the quantum systems in each single
run of the CKS protocol, the values of $x_{0}^{(i)}$, $x_{1}^{(i)}$ in
different runs will not correlate with each other. Then in the current case,
since $x_{0}^{(i)}$, $x_{1}^{(i)}$ in all the $n$ runs are kept at the
quantum level, any one of them can turn out to be either $0$ or $1$ during
the measurement in the security check. The outcome is determined
independently in each single run by quantum uncertainty, thus Bob cannot
ensure with probability $100\%$\ that $X_{0}$ and $X_{1}$ will always take
the legitimate values in set $S$ in every single run in the security check.
Therefore, we have $p_{c}<1$ when $p=1$. Then equation (\ref{pa}) gives $%
P_{Alice}^{\ast }=(1/2+1/4)p_{c}<3/4$. Namely, no matter $p<1$ or $p=1$, $%
P_{Alice}^{\ast }$ cannot equal exactly to $3/4$, i.e., it is always lower
than that of the original CKS protocol.

Meanwhile, Bob's cheating probability still equals to that of the CKS
protocol, i.e., $P_{Bob}^{\ast }=1/2$, since the specific $x_{0}^{(i_{3})}$,
$x_{1}^{(i_{3})}$\ finally chosen for encoding the target bits $x_{0}$, $%
x_{1}$ are not affected by any $x_{0}^{(i)}$, $x_{1}^{(i)}$ from all the
other runs, as it is elaborated in the previous subsection. Combining this $%
P_{Bob}^{\ast }$\ with $P_{Alice}^{\ast }<3/4$, we can see that under the
limited collective attack, our Protocol B can obtain%
\begin{equation}
2P_{Alice}^{\ast }+P_{Bob}^{\ast }<2\times \frac{3}{4}+\frac{1}{2}.
\label{protocol B}
\end{equation}%
Thus the upper bound $2P_{Alice}^{\ast }+P_{Bob}^{\ast }<2$\ is proven.

Now we prove the lower bound $2P_{Alice}^{\ast }+P_{Bob}^{\ast }\geq 5/3$.
This is because there exists the following cheating strategy for Alice. In
every triple run, she only keeps one pair of $x_{0}^{(i)}$, $x_{1}^{(i)}$ ($%
i=1,2,3$) at the quantum level by using the collective operation described
by equation (\ref{T}). The other two pairs of $x_{0}^{(i)}$, $x_{1}^{(i)}$
are all taken as $0$ beforehand, and the corresponding two runs of the CKS
protocol are executed honestly. Then all triple runs can pass the security
check with certainty. Meanwhile, when a useful run is finally picked for
encoding the target bits, the pair $x_{0}^{(i)}$, $x_{1}^{(i)}$ kept at the
quantum level stands a probability $1/3$ to be chosen. Thus we have $p=1/3$.
Substitute it into equation (\ref{pa}) and we yield%
\begin{equation}
2P_{Alice}^{\ast }+P_{Bob}^{\ast }=\frac{5}{3},  \label{lower}
\end{equation}%
so that this lower bound can be reached even when Alice is restricted to the
limited collective attack. But we do not know whether this bound is tight at
the present moment, as it is unclear whether there may exist an even better
cheating strategy.

\section{Potential improvements}

By observing the above cheating strategy that led to the lower bound
equation (\ref{lower}), we can see that the probability $p=1/3$ comes from
the specific set $S$ used in Protocol B, which is made of $3$-bit strings
only. In the more general form, i.e., our Protocol A, we can expect that
choosing a more complicated set $S$ may further reduced the value of $p$.
For example, set $S$ can be chosen as a classical error-correcting code,
e.g., the binary linear $(n,k,d)$-code \cite{qi43}. That is, $S$ is taken as
a set of classical $n$-bit strings. Each string is called a codeword. This
set of strings has two features. (a) Among all the $2^{n}$ possible choices
of $n$-bit strings, only a particular set of the size $\sim 2^{k}$\ ($k<n$)
is selected to construct this set. (b) The distance (i.e., the number of
different bits) between any two codewords in this set is not less than $d$\ (%
$d<n$). With these features, it can be expected that by increasing $n$ while
fixing $k/n$ and $d/n$, a dishonest Alice will have to introduce a much
larger number of entangled control qubits for keeping more pairs of $%
x_{0}^{(i)}$, $x_{1}^{(i)}$ at the quantum level, so that $%
X_{0}=x_{0}^{(1)}x_{0}^{(2)}...x_{0}^{(i)}...x_{0}^{(n)}$ and $%
X_{1}=x_{1}^{(1)}x_{1}^{(2)}...x_{1}^{(i)}...x_{1}^{(n)}$ will appear as
legitimate strings in set $S$ no matter which bits are picked for the
security check. Therefore with a properly chosen $S$, Protocol A may further
lower the successful probability of limited collective attacks, and also
raises the difficulty of implementing these attacks. However, the rigorous
security bound will depend heavily on the structure of the specific set $S$
used in the protocol. This analysis is left for future research.

\section{Summary and remarks}

Thus we show that the security bound $2P_{Alice}^{\ast }+P_{Bob}^{\ast }\geq
2$ for weak OT can be violated for an Alice with limited computational
power. As a rigorously checkable example, we proposed Protocol B which
reaches the maximal violation $2P_{Alice}^{\ast }+P_{Bob}^{\ast }\rightarrow
3/2$ when only individual measurements are allowed. For attacks using
collective operations on a limited number of quantum systems, there can
still be $P_{Alice}^{\ast }<3/4$ while $P_{Bob}^{\ast }=1/2$. An even lower
value of $P_{Alice}^{\ast }$ could be expected from Protocol A.

Note that Ref. \cite{qbc88} obtained the security bound without limiting
Alice to individual measurements. Thus our protocols does not really break
the bound in principle. But it still has great practical significance. This
is because in practice, any quantum storage devices can keep the quantum
states faithfully for a limited time only. Suppose that $\tau $ is the
maximal storage time available with state-of-the-art technology. Then during
step B1 of Protocol B, Bob can require that every run of the CKS protocol is
separated from each other by a time interval larger than $\tau $, so that
any ancillary system that a dishonest Alice may introduce to entangle with
Bob's qutrit will suffer from errors, making Alice unable to pass the
security check. In this case, Alice has to finish the measurement on Bob's
qutrit (if she does not want to perform the honest unitary transformation)
in each single run of the CKS protocol before the next run begins. Thus her
cheating is actually reduced to individual measurements. So we can see that
as long as our protocol is proven secure against individual measurements,
then it naturally implies that we can use it as a secure protocol in
practice.


This work was supported in part by
the NSF of Guangdong province.


\begin{thebibliography}{99}
\bibitem{qbc9} Rabin, M. O.: How to exchange secrets by oblivious transfer.
Technical Report TR-81, Aiken Computation Laboratory, Harvard University
(1981) (available online at http://eprint.iacr.org/2005/187.pdf)

\bibitem{1-2OT} Even, S., Goldreich, O., Lempel, A.: A randomized protocol
for signing contracts. In: Advances in Cryptology: Proc. Crypto '82, ed.
Chaum, D., Rivest, R. L., Sherman, A. T., Plenum (1982), p. 205

\bibitem{qi139} Kilian, J.: Founding crytpography on oblivious transfer. In:
Proc. 1988 ACM Annual Symposium on Theory of Computing, ACM, New York
(1988), p. 20 

\bibitem{qi499} Colbeck, R.: Impossibility of secure two-party classical
computation. Phys. Rev. A \textbf{76}, 062308 (2007)

\bibitem{qi677} Salvail, L., Schaffner, C., Sotakova, M.: On the power of
two-party quantum cryptography. e-print. arXiv:0902.4036 (2009)

\bibitem{qi725} Salvail, L., Sotakova, M.: Two-party quantum protocols do
not compose securely against honest-but-curious adversaries. e-print.
arXiv:0906.1671 (2009)

\bibitem{qi797} Colbeck, R.: Quantum and relativistic protocols for secure
multi-party computation. e-print. arXiv:0911.3814 (2009)

\bibitem{qbc14} Chailloux, A., Kerenidis, I., Sikora, J.: Lower bounds for
quantum oblivious transfer. Quantum Inf. Comput. \textbf{13}, 158 (2013)

\bibitem{qbc88} Chailloux, A., Gutoski, G., Sikora, J.: Optimal bounds for
quantum weak oblivious transfer. e-print. arXiv:1310.3262 (2013)

\bibitem{qi51} Bennett, C. H., Brassard, G., Cr\'{e}peau, C., Skubiszewska,
M.-H.: Practical quantum oblivious transfer. In: Advances in Cryptology:
CRYPTO '91, Lecture Notes in Computer Science, vol. \textbf{576}, ed.
Feigenbaum, J., Springer-Verlag (1992) p. 351

\bibitem{qi138} Mayers, D., Salvail, L.: Quantum oblivious transfer is
secure against all individual measurements. In: Proc. Third Workshop on
Physics and Computation - PhysComp '94, IEEE Computer Society Press, Dallas
(1994), p. 69

\bibitem{qi24} Mayers, D.: Unconditionally secure quantum bit commitment is
impossible. Phys. Rev. Lett. \textbf{78}, 3414 (1997)

\bibitem{qi23} Lo, H.-K., Chau, H. F.: Is quantum bit commitment really
possible? Phys. Rev. Lett. \textbf{78}, 3410 (1997)

\bibitem{qi140} Cr\'{e}peau. C.: Equivalence between two flavours of
oblivious transfers (abstract). In: Advances in Cryptology: CRYPTO '87,
Lecture Notes in Computer Science, vol. \textbf{293}, ed. Pomerance, C.,
Springer-Verlag (1988), p. 350

\bibitem{HeJPA} He, G. P.: Quantum key distribution based on orthogonal
states allows secure quantum bit commitment.\ J. Phys. A: Math. Theor.
\textbf{44}, 445305 (2011)

\bibitem{HeQOT} He, G. P., Wang, Z. D.: Oblivious transfer using quantum
entanglement. Phys. Rev. A \textbf{73}, 012331 (2006)

\bibitem{qi135} Shimizu, K., Imoto, N.: Communication channels analogous to
one out of two oblivious transfers based on quantum uncertainty. Phys. Rev.
A \textbf{66}, 052316 (2002)

\bibitem{qi43} Brassard, G., Cr\'{e}peau, C., Jozsa, R., Langlois, D.: A
quantum bit commitment scheme provably unbreakable by both parties. In:
Proc. 34th Annual IEEE Symposium on Foundations of Computer Science, IEEE,
Los Alamitos (1993), p. 362
\end{thebibliography}
\end{document}